\documentclass[12pt]{article}
\usepackage[dvips]{graphicx}

\def\Vec#1{{\mbox{\boldmath $#1$}}}
\def\beqa{\begin{eqnarray}}
\def\eeqa{\end{eqnarray}}
\def\beq{\begin{equation}}
\def\eeq{\end{equation}}
\def\fracp#1#2{\frac{\partial#1}{\partial#2}}
\def\non{\nonumber}
\def\L{\left}
\def\R{\right}
\def\ss{\Vec{\sigma}}
\def\Tr{\mbox{Tr}}

\def\NPB#1#2#3{Nucl. Phys. {\bf B#1}(#2)#3}
\def\PLB#1#2#3{Phys. Lett. {\bf B#1}(#2)#3}

\makeatletter
\@addtoreset{equation}{section}

\begin{document}
\renewcommand{\thefootnote}{\fnsymbol{footnote}}
\begin{titlepage}
\begin{flushright}
hep-th/0307058v3\\ %%%%%%%%%%%%%%%%%%%%%%%%%%%v3%%%%%%%%%%%%%%%
UT-KOMABA/03-11\\
April 2004 %%%%%%%%%%%%%%%%%%%%%%%%%v3%%%%%%%%%%%%%%%%%%%
\end{flushright}
\begin{center}

{\bf\Large Membrane topology and matrix regularization}

\vspace{1.5cm}

{\bf  Hidehiko~Shimada}\footnote
{
E-mail: shimada@hep1.c.u-tokyo.ac.jp
}	

\vspace{1.0cm}

{\it  Institute of Physics, Tokyo University,\\
Komaba, Megro-ku, Tokyo 153-8902, Japan}

\vspace{1.9cm}
\end{center}
\begin{abstract}
The problem of membrane topology in the matrix model of M-theory is considered. 
The matrix regularization
procedure, which makes a correspondence  between finite-sized matrices
and functions defined on a two-dimensional base space,
is reexamined.
It is found that the information of topology of the base space
 manifests itself in the eigenvalue distribution of a single matrix.
The precise manner of the manifestation is described.
The set of all eigenvalues can be decomposed into 
subsets whose members increase smoothly, provided that the 
fundamental approximations in matrix regularization hold well.
Those subsets are termed as eigenvalue sequences.
The eigenvalue sequences exhibit a branching phenomenon
which reflects Morse-theoretic information of topology.

Furthermore, exploiting the notion of eigenvalue sequences,
a new correspondence rule
between matrices and functions is constructed.
The new rule identifies the matrix elements directly with
Fourier components
of the corresponding function, evaluated along certain orbits.
The rule has semi-locality in the base space, so that it can be used
for all membrane topologies in a unified way.
A few numerical examples are studied, and  
consistency with previously known correspondence rules is discussed.
\end{abstract}
\vfill
\end{titlepage}
\setcounter{footnote}{0}
\renewcommand{\thefootnote}{\arabic{footnote}}

\section{Introduction}
Topological properties of a system are often important in 
investigating the dynamics of the system.
It seems certain that
M-theory\cite{11_10} 
has membranes as dynamical degrees of freedom.
Furthermore, the only existing proposal for formulation of M-theory, 
namely the matrix model of M-theory\cite{IMF}, can be considered
as an attempt to define quantum membrane
theory~\cite{Memb_Mr_mrG, Memb_Mr_mrH, S_Memb_Mr}.
More explicitly stated, it is a regularized version of 
membrane theory in lightcone gauge, dynamical
variables becoming $N \times N$ matrices instead of functions
defined on two-dimensional worldspace.

But, at present, the topological properties of membranes in M-theory are not 
known.
The concern of this paper is membrane topology
\footnote{
We use the word membrane topology 
to express the topology of a configuration of membranes in a time-slice.
The topology is not of a single membrane but of
a totality of membranes.
Thus, membrane topology is classified by the numbers of
membranes $n_i$ which has genus $i=0,1,\cdots$.
}
in the matrix model.
It is believed that the matrix model can describe
membranes of arbitrary topologies.
However, there has been a problem:
we do not know 
whether and how the information of the 
topology manifests itself in the matrix model.
The cause for this problem lies in the manner
in which the correspondence between matrices and functions has been given.
There has been no unique rule that can deal with all membrane topologies.
Instead, we have many different rules for different topologies
\cite{Memb_Mr_mrG, Memb_Mr_mrH, S_Memb_Mr, MrTor, MrGen},
interrelationships between those rules being unclear.

In this paper, we address this problem
by reexamining the regularization procedure,
the so-called matrix regularization. 
We shall show that the information indeed manifests itself 
in the eigenvalue distribution of a single matrix.
The precise manner of the manifestation will be described.
Moreover, we have constructed
 a new correspondence rule between functions and matrices
which can be applied to all membrane topologies in a unified way.

We start by discussing 
relevant aspects of matrix regularization, in section \ref{SMatReg}.
We take the simple view that, classically,
the matrix regularization is an approximation
of continuum theory by a discretized theory.
\footnote{
In quantum theory, at the same time, 
the matrix regularization is considered as
 a definition of continuum
theory by a non-trivial limit of discrete theories.
This is the reason why we should first treat the finite-$N$
theory carefully.
}
This approximation between two theories
is based solely on some fundamental large-$N$ approximation
formulae (\ref{AppNrm})-(\ref{AppBra}).
They play a vital role in this paper.
We also recall the
well-known mathematical analogy between
the matrix regularization and 
canonical quantization
of systems with one degree of freedom, which will be our main tool in 
subsequent discussions.

Then, in section \ref{STop}, 
we turn to the investigation of membrane topology 
in the matrix model.
Our basic observation is that, in order to study membrane topology,
it suffices to consider the two-dimensional base space, which we 
shall term as the $\ss$-space,
not the shape of the membranes in the target space. 
This observation greatly simplifies the analysis,
since it enables us to deal with only a single matrix, not many matrices.

We base the discussion on the analog of the Bohr-Sommerfeld quantization
condition.
We shall show that,
in the case where the fundamental approximations hold well,
the eigenvalue distribution of a matrix has a particular structure.
Namely, the set of all eigenvalues can be decomposed into subsets 
characterized by the following property:
the eigenvalues in one of the subsets, when sorted, increase smoothly.
We call these subsets as eigenvalue sequences.
The grouping of the eigenvalues into sequences
reveals a branching phenomenon of sequences.
We find that the branching phenomenon, in turn, reflects 
certain Morse-theoretic information of topology of the $\ss$-space. 
This is our answer to the above problem. Thus, the information of topology
manifests itself, in the  world of matrices, as a branching phenomenon of 
eigenvalue sequences.

Furthermore, the notion of eigenvalue sequences enables us to construct
a new correspondence rule between matrices 
and functions, which is the subject in section \ref{SMRCorr}. 
The matrix elements are approximately equal to 
Fourier components of the corresponding function,
calculated along appropriate orbits on the $\ss$-space.
The rule is analogous to the correspondence 
noticed by Heisenberg when he created Matrix Mechanics
pursuing Bohr's correspondence principle\cite{Heisenberg}.
There, the matrix elements of an observable, 
in the basis which makes
the Hamiltonian diagonal, are equal to the classical
Fourier components of the observable along the appropriate
classical orbits on the phase space. 
We shall show that the fundamental approximation formulae
hold well if the new correspondence rule
holds.
The correspondence rule contains the above-mentioned analog of the
Bohr-Sommerfeld condition.
This justifies the use of it 
in section \ref{STop}.

The new rule is  semi-local in the $\ss$-space, and consequently
can be applied for all membrane topologies uniformly, in
marked contrast with the previously known rules.
This, in particular, enables one to construct functions corresponding
to given matrices when the approximations are good. Using previous rules,
one could only do the reverse, namely, to construct matrices corresponding 
to given functions.
This is because one could not know the topology corresponding to
the given matrices, and therefore could not choose the rule to be used.

Apart from the unified treatment for all topologies, the new rule 
has the virtue that the identification of the matrix elements 
with Fourier components is direct, and
so that the geometrical meanings of the
matrix elements
are clear.
Our arguments are 
also relevant to the matrix model of type IIB string theory
\cite{TypeIIBMatrixModel},
since the same regularization is involved.
Further, the same kind of mathematics as that of matrix regularization appears
in such subjects as bound states of D-branes or 
non-commutative field theory. Ideas in this paper may find some applications
in those subjects.

A few illustrative numerical examples are given
in section \ref{SNumEx}.
The consistency between our new rule and previous rules
is checked by studying them. Finally, we conclude with some discussions
in section \ref{SConDis}.

\section{Matrix regularization} \label{SMatReg}
Let us briefly recall the matrix regularization procedure
from our viewpoint.
Although it is supermembrane theory in eleven dimension
\cite{S_Memb_Action}
that is relevant to M-theory,
we consider  bosonic membrane theory 
for simplicity of presentation.  

Firstly, we shall describe the continuum theory.
We parametrize the 
membranes by three parameters $(\tau,\sigma^1,\sigma^2) =
(\tau,\ss)$.
Then, the geometrical shape of membranes in spacetime is described by the coordinate
functions
$x^\mu(\tau,\ss)$. 
In lightcone gauge formalism,
$\tau$ is chosen to be equal to $x^+$, and $\ss$ is chosen so as to
make the area of a domain in the $\ss$-space proportional to  total 
$p^+$ contained in the domain. Here, we denote 
the momentum density vector of the membranes by $p^\mu$.
The canonical variables of the system are
transverse coordinates and momenta (which are functions 
defined on the $\ss$-space)
as well as zero modes, 
\beq
x^\alpha(\ss), p^\alpha(\ss); X^-, -P^+.\label{ContiCanVar}
\eeq
The Hamiltonian is given by 
\beq
H=[\ss]\int\frac{(p^\alpha)^2+
\frac{1}{2}(\{x^\alpha,x^\beta\})^2}{2 P^+} d^2\ss,
\label{ContiH}
\eeq
with Lie brackets 
\[
\{ f ,g \}= \fracp{f}{\sigma^1}\fracp{g}{\sigma^2}-
\fracp{f}{\sigma^2}\fracp{g}{\sigma^1}, \non
\]
where $f$ and $g$ are functions on the $\ss$-space. 
We have also introduced a conventional constant
$[\ss]$ which is the total area of the $\ss$-space,
$[\ss] =\int d^2\ss$.
The remaining ingredients of the theory are the phase space constraints
\beq
\{x^\alpha, p^\alpha\}(\ss)=0,\label{ContiCon}
\eeq
and its global version. They correspond to the local symmetry of the 
lightcone gauge theory
under reparametrization by area-preserving diffeomorphism (APD) on 
the $\ss$-space.
This is a local symmetry, because one can perform reparametrization 
by different 
APD for different $\tau$.

Secondly, we shall give the regularized theory.
The canonical variables are
$N\times N$ matrices as well as  zero modes,
\beq
\hat{x}^\alpha,\hat{p}^\alpha; X^-, -P^+.\label{RegCanVar}
\eeq
The Hamiltonian is given by
\beq
H= N\Tr \frac{(\hat{p}^\alpha)^2-\frac{1}{2}(2\pi
[\hat{x}^\alpha,\hat{x}^\beta])^2}
{2 P^+},\label{RegH}
\eeq
and the constraints are,
\beq
[\hat{x}^\alpha,\hat{p}^\alpha]=0,\label{RegCon}
\eeq
where $[~,~]$ is a commutator of matrices.

Now, we turn to the explanation of 
the matrix regularization.
The following fact is essential:
there exists a correspondence between appropriate
functions on the $\ss$-space
$f(\ss),g(\ss),\cdots$ 
and matrices $\hat{f}, \hat{g}, \cdots$ such that
the fundamental approximation 
formulae
\beqa
\frac{1}{[\ss]}\int f(\ss) d^2 \ss &\approx& \frac{1}{N} \Tr \hat{f} 
\label{AppNrm}\\
\widehat{f g} &\approx& \hat{f} \hat{g}
\label{AppMtp}\\
\widehat{\{f,g\}} &\approx& -i \frac{2\pi N}{[\ss]} [ \hat{f} ,\hat{g} ]
\label{AppBra}
\eeqa
hold.
Here, we denote by $\widehat{\{f,g\}}$ and $\widehat{f g}$
the matrices which correspond to the functions $\{f,g\}(\ss)$
and $f(\ss) g(\ss)$, respectively.
\footnote{
Maybe we should add the linearity of the correspondence,
$\widehat{f+g}=\hat{f}+\hat{g}$,
for the sake of completeness.
We have omitted it since it holds trivially in all our discussions.
}
The larger is $N$, the better is the approximation.
From these formulae it follows that 
the continuum theory, defined by (\ref{ContiCanVar})-(\ref{ContiCon})
can 
be approximated by a regularized theory defined by
(\ref{RegCanVar})-(\ref{RegCon}).
We stress the importance of above formulae.
They are almost the definition of the matrix regularization.

Since Lie brackets and matrix commutators 
both obey the Jacobi identity and antisymmetry, the important 
advantage of matrix regularization follows.
Namely, the regularized theory has local symmetry under the transformation
\beq
x^{\alpha\prime}(\tau)=U(\tau)x^\alpha(\tau) U(\tau)^{-1}, \ \ 
p^{\alpha\prime}(\tau)=U(\tau)p^\alpha(\tau) U(\tau)^{-1}, \label{LocSymMat}
\eeq
where $U(\tau)$ is an arbitrary matrix which is a function of $\tau$,
corresponding to
the APD symmetry in continuum theory.

The matrix regularization procedure is analogous to the
quantization of a system which has one degree of freedom,
as is well known. The analogy can be summarized as,
\beq
\begin{array}{c|c}
\mbox{Canonical quantization} 
& \mbox{Matrix regularization} \\
\hline
 (x,p)& (\ss^1,\ss^2)\\
 \mbox{Canonical transformation}& \mbox{Area-preserving diffeomorphism}\\
\{~,~\}_{P.B.} \rightarrow -i\frac{1}{\hbar} [~,~]&
\{~,~\} \rightarrow -i\frac{2\pi N }{[\ss] } [~,~]\\
\hbar & \frac{[\ss]}{2\pi N} \\
\end{array} \label{AnalogyTable}
\eeq
where $\{~,~\}_{P.B.}$ is the usual Poisson brackets.
We shall motivate our discussion by this analogy in section \ref{STop}.

We conclude this section with discussions on 
the previously known correspondence rules. 
We first recall the general manner the rules are formulated. 
We must, 
first of all, fix topology of the $\ss$-space.
After that, we consider a basis in the vector space of all functions 
defined on the $\ss$-space.
Then, we define an appropriate basis in the vector space of all 
$N \times N$ matrices, and postulate a correspondence
between it and
the basis in the space of the functions appropriately truncated.
The rules are, finally,
justified by checking
that the fundamental approximations (\ref{AppNrm})-(\ref{AppBra}) 
hold well for large $N$ by them.

This manner has made difficult
to consider whether and how membrane topology
manifests itself in the matrix model.
In particular,   
one can expand an arbitrary matrix 
by basis referring to any particular topology.
This fact, at first sight, seems to suggest 
that a configuration of matrix model 
could  be interpreted as membranes of arbitrary topology,
and there would be, therefore, no information of topology
in the matrix model.

This is not necessarily true. 
Even if one can formally expand some matrices
by a basis referring to a particular topology,
the fundamental approximations  may not work at all.
\footnote
{
This may be expected from the previously known rules.
Let us, for example, imagine a smooth function defined 
on a torus. One can construct the corresponding matrix
using the basis for torus topology.
One can then expand the matrix by the basis (in the space of matrices)
corresponding to the topology of a sphere, and construct a function
defined on a sphere. We expect that the resulting function
would have discontinuity or, in any case, some singularity
(see subsection II. C of \cite{ReviewByTaylor}).
This implies that 
the function varies considerably in a very small length scale.
Therefore the approximations may well be no good,
since, in general, 
the smaller the length scale of the variation of functions,
the larger must be $N$ in order that the approximations are good.

However, it is difficult to characterize precisely, 
using only previously known correspondence rules, when 
the approximations break down. 
Hence,
it has not been clear if this picture is indeed right.
}
In our perspective, that the matrix regularization is an approximation
scheme, we cannot, then, interpret the matrices as membranes of the
particular topology.
Information of topology 
may be hidden in the matrices in this way.
Through sections \ref{STop} and \ref{SMRCorr},
we shall see indeed that, 
provided that the approximations are good,
the information reflects in the
eigenvalue distribution.

\section{Membrane topology and matrix regularization}
\label{STop}
In this section, we show that the information of membrane topology
manifests itself in the matrix model.
Before explicit description of the manner of the manifestation,
let us give some basic observations.

If one wishes to specify the complete shape of membranes in the
 target space,
one needs information of many (that is, roughly speaking, 
as many as the dimension of the target space)
functions.
This would imply that one should study many matrices in the matrix model.
However, the information of membrane topology,
or at least the information of 
topology of the $\ss$-space,
is contained in one generic 
function defined on the $\ss$-space,
as is strongly suggested by Morse theory.
We choose, as our basic strategy, to consider the latter information.
Then, we shall seek  in a single 
 matrix the information of topology of the $\ss$-space.

There is another point we would like to discuss.
It is most natural to identify functions which are transformed into each
other by APD transformations.
We shall identify those matrices which are transformed into
each other by similarity transformations, 
since (\ref{AppBra}) tells us that
the counterpart of the APD transformation
is the similarity transformation.
This has some non-triviality, since 
it may happen that the identification is only allowed
approximately. Nonetheless, we shall carry out the identification,
because that the APD symmetry survives as (\ref{LocSymMat}) is the most
important advantage of the matrix regularization.
This identification and our strategy, to consider 
the topology of the $\ss$-space, act together to
greatly simplify the analysis.
Since one can always diagonalize a single matrix,
we can concentrate on the eigenvalue 
distribution of the matrix.

Having explained our basic strategy,
we shall now proceed to investigate the manner of the manifestation
of membrane topology in the eigenvalue distribution.

First, let us consider how one can read off the information of topology
from a function in an APD invariant way.
We choose an arbitrary generic function $f(\ss)$.
\footnote
{
We use the word generic in the sense of Morse theory:
we avoid  degenerate functions, constant functions for instance, 
which can be changed into  generic functions by arbitrarily small perturbations.
}
It could be one of the transverse coordinates, for instance.
The function is fixed, throughout our discussion, 
as a kind of reference.
Thus, we shall use $f$, shortly below, as both
an analog of the Hamiltonian in canonical 
quantization and a Morse function.
As a natural APD-invariant concept with a given function $f(\ss)$,
we introduce
an ordinary differential equation (ODE)
\beq
\frac{d}{dt}\ss =\{ \ss,f\} \label{EomOnSs}, 
\eeq
drawing analogy to the Hamiltonian equation 
of motion with a given Hamiltonian function $H(x,p)$, 
which is invariant under canonical transformations.
This ODE governs the motion of points of the $\ss$-space.
\footnote
{
We note that the independent variable of the ODE, $t$, is
just a mathematical tool to substantiate the analogy to canonical
quantization. It has nothing to do with physical time coordinates
of membrane theory.
}
Thus, we envisage an auxiliary Hamiltonian-like
dynamical system with the $\ss$-space as its phase space
and with $f$ as its Hamiltonian.
Since $f$ is conserved along the motion by the identity $\{f, f\}=0$,
an orbit of this equation is a part of an equal-$f$ line in the $\ss$-space.
It will form a closed loop because of the compactness of the $\ss$-space.
%%%%%%%%%%%%%%%%%%%%%%Ftorus%%%%%%%%%%%%%%%%%%%
\begin{figure}[t]
\begin{center}
\includegraphics[width=.4\linewidth]{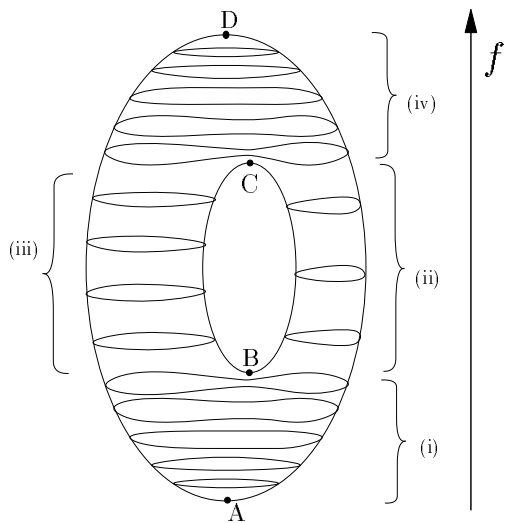}
\caption {The $\ss$-space which has topology of a torus.
The height in the figure is the reference function.
Some orbits of the ODE $d \ss/dt=\{ \ss, f\}$ are drawn,
which form closed loops that are, in turn, parts of
the equal-$f$ lines.
If one gradually increases the value of $f$,
one observes a branching phenomenon of the orbits:
appearing, branching, merging, and disappearing processes at the points
A, B, C, and D, respectively.
Depicted is essentially the $\ss$-space, 
so that the horizontal directions of the figure
have rather arbitrary meanings.
If one wishes, one can also give definite meanings to the
horizontal directions by interpreting 
this figure as the geometrical shape
of a membrane in the target space, and the reference function
as one of the coordinate functions. 
} 
\label{Ftorus}
\end{center}
\end{figure}
%%%%%%%Ftorus%%%%%%%%%%%%%%%%%%%%%%%%%%%%%%%%%%%%%%%%%%%%%

Then, 
if we scan the $\ss$-space by gradually increasing the value of $f$,
we will observe 
branching processes of these orbits.
There are four types of these branching processes:
appearing, disappearing, branching  and merging.
Let us consider, for a typical example, the situation depicted 
in Fig.~\ref{Ftorus}.
The membrane topology is that of a torus.
The reference function $f$
is chosen to be the height in the figure, and 
some orbits of (\ref{EomOnSs}) are drawn.
In this example, at the points A, B, C, D, 
the orbits appear, branch, merge, disappear, respectively.
We can read off the information of topology from these processes.
This is just the well-known idea of 
Morse theory.
In particular, we obtain the Euler number of the $\ss$-space,
by subtracting the total number of the 
branching and  merging processes from
the total number of the appearing and disappearing processes.

Now, we shall show that this analysis of topology in the world of 
functions has a counterpart in the world of matrices. 
The analogy of the matrix regularization
to  canonical quantization is useful here.
In the latter, the Bohr-Sommerfeld quantization condition
determines the eigenvalues of the Hamiltonian operator $\hat{H}$ from the classical Hamiltonian function $H(x, p)$ defined on the phase space $(x, p)$.
Namely, we draw classical orbits in $(x, p)$ space, 
that are parts of equal-$H$ lines,
so that the areas of the domains between two adjacent orbits are equal to
$2\pi \hbar$. Then eigenvalues of $\hat{H}$ are given by 
the values of $H$ at these orbits.
Here, we shall
exploit the analogy, which is summarized in (\ref{AnalogyTable}),
and state the analog of the Bohr-Sommerfeld condition.
Namely, we draw orbits of (\ref{EomOnSs}) in the $\ss$-space
so that the areas of the domains
\footnote
{
The area of a domain means here area in the $\ss$-space
not in the target space.
Its physical meaning is the total $p^+$ contained in the domain,
apart from a conventional factor,
by the gauge choice made in the lightcone gauge formalism.
}
 between two adjacent orbits are equal to
$[\ss]/N$.
Since $[\ss]$ is the total area of the $\ss$-space,
this simply means that we divide the $\ss$-space
into $N$ parts of equal area.
Eigenvalues of $\hat{f}$ are then given by 
the values of $f$ at these orbits. 
We assume this rule to hold. We shall justify the assumption
in section \ref{SMRCorr}.

If we apply this rule to the case in Fig.~\ref{Ftorus}, then 
the eigenvalues of $\hat{f}$ can be grouped into 
four subsets each of which corresponds to the family of
the orbits belonging to 
(i) the region from the point A to 
the point B, (ii) the left branch of the torus from the point B 
to the point C,
(iii) the right branch of the torus from the point B to the point C,
(iv) the region  from the point C to the point D, respectively.
We call these subsets as 
eigenvalue sequences.
For large enough N, eigenvalues belonging to each sequence
 have the following property.
If we sort the eigenvalues contained in a sequence 
in increasing order of their values, and make a graph 
plotting the values of them versus 
their order,
then the plotted points can be linearly approximated locally.
To put it short, 
the eigenvalues in a sequence
increase smoothly.
It should be clear that, in general, if we do not group the eigenvalues 
properly,
then the graph become zigzag-shaped and the above property is  lost.
In section \ref{SMRCorr}, we see that
this linear approximation is  essential in order 
 the fundamental approximations
(\ref{AppNrm})-(\ref{AppBra}) to hold.

The eigenvalue sequences should exhibit the same branching phenomenon
as that of the orbits.
For the example of Fig.~\ref{Ftorus}, the sequence
(i) appears and then branches into the sequences (ii) and (iii).
They merge into the sequence (iv), and finally (iv) disappears.
It is clear that all these considerations work the same in general cases
other than that of Fig.~\ref{Ftorus}.
Thus, the information of membrane topology manifests itself
in the branching phenomenon (which consists of appearing, branching, merging,
disappearing processes) of eigenvalue sequences.
A few examples, including the case similar to the situation in 
Fig.~\ref{Ftorus}, are given in section \ref{SNumEx}.

\section{The new correspondence rule} \label{SMRCorr}
In this section, we present a new 
correspondence rule between matrices and functions,
and then show that the fundamental approximations
(\ref{AppNrm})-(\ref{AppBra})
stem from the rule.

We choose an arbitrary
 generic function $f$ and fix it as a reference, as in section 
\ref{STop}.
The rule is formulated in such a way that
the representation of matrices is so chosen that
 the matrix $\hat{f}$, corresponding to 
the function $f$, is diagonal.
For simplicity of notation,
we shall consider the case where only one
eigenvalue sequence is present. We explain 
the generalization later in this section. 

We first give the rule to determine the diagonal
matrix $\hat{f}$.
To this end, we set up some notations. 
We again consider 
ODE (\ref{EomOnSs})
\[
\frac{d\ss}{d t}= \{ \ss, f \} .
\]
A solution of this ODE is periodic, the point of the $\ss$-space 
circulating on a loop which is part of an equal-$f$ line. 
We shall denote its period, as a function of $f$, by $T(f)$.
We sort the eigenvalues of $\hat{f}$ 
in increasing order, and call them $\hat{f}_n$,
\beq
\cdots \le \hat{f}_{n-1} \le \hat{f}_n \le \hat{f}_{n+1} \le \cdots. 
\eeq
To be specific, we choose the representation such that
\beq
\hat{f}=\mbox{diag}(
\cdots, \hat{f}_{n-1}, \hat{f}_n, \hat{f}_{n+1}, \cdots 
).
\eeq
The relation between the function $f$ and the 
matrix elements $\hat{f}_n$
is the analog of the Bohr-Sommerfeld quantization condition stated in 
section \ref{STop}. If $N$ is sufficiently large, the rule can be 
formulated as,
\beq 
\hat{f}_m-\hat{f}_n \approx (m-n) \frac{[\ss]}{ N}
 \frac{1}{T\bigl(\!
 \frac{\hat{f}_m+\hat{f}_n}{2}\!\bigr)}, \label{QtzOfArea}
\eeq 
when $|m-n|$ is small. 
We have used that for two nearby loops, one at 
$f$ and the other at $f+\delta f$,
the area $\delta S$ between them can be approximated by
\beq
\delta S=\oint \frac{\delta f}{|
\mbox{grad} f|}ds=T\!\biggl(\!f+\frac{\delta f}{2}\biggr)\ \delta f .
\eeq
We can construct $\hat{f}_m$ satisfying (\ref{QtzOfArea}) directly 
by the following method. We first define $S(f) = \int^f (1/T(f)) df$.
The value of $S(f)$ runs from $0$ to $[\ss]$ in this case
where there is
only one eigenvalue sequence.
We then consider the inverse function $f(S)$, and set
$\hat{f}_m=f(S_m)$, where $S_m$ are determined 
by $S_{m+1}-S_m= [\ss]/N$ up to a constant shift.
The shift should be of order $1/N$ for consistency.
\footnote 
{
We can determine the shift  by setting $S_1=[\ss]/(2N)$
for an eigenvalue sequence beginning
with an appearing process. 
%This means 
%the area of
%the part of the $\ss$-space satisfying
%$f(\ss)\leq \hat{f}_1$ is equal to $[\ss]/(2N)$. 
This is analogous to 
the $1/2$ in the Bohr-Sommerfeld condition 
$\oint p dq = (n + (1/2) ) 2 \pi \hbar$.
%, representing the effect of 
%zero point energy.
The justification for the above rule comes from
the fact that
(\ref{DerivedAppNrm}) holds at one more higher order
in $1/N$ by this rule.
In other words, the rule is just 
the midpoint rule for numerical integration.
Similar rule exists for an eigenvalue sequence
ending with a disappearing process.
\label{FootnoteOnShiftZeroPoint}}

%%%%%%%%%%%%%%%%%%%%% a loop integral along these loops

Having stated the correspondence rule for the reference function,
we next turn to the correspondence rule for an arbitrary function $g$.
We denote the matrix elements of the corresponding matrix $\hat{g}$
by $\hat{g}_{mn}$.
When $|m-n|$ is small, $\hat{g}_{mn}$ is equal to the 
Fourier component of
order $m-n$ of the function $g(\ss(t))$.
Here, $\ss(t)$ denotes the solution of (\ref{EomOnSs}) along which
the function $f(\ss)$ takes the (constant) value $(\hat{f}_m+\hat{f}_n)/2$. 
To obtain explicit formulae,
we define the Fourier components $g_s(f)$ by
\beq
g(\ss(t))=\sum_{s=-\infty}^{+\infty}
g_s(f)\ e^{i \left(\frac{2\pi}{T(f)} s \right) t}, \label{DefOfgOfs}
\eeq
where the parameter $f$ denotes the value of the function $f(\ss)$
along the solution $\ss(t)$.
We then set,
\footnote{
This relation is the direct
analog of the correspondence, in semi-classical region,
between quantum matrix elements
and Fourier components
along classical orbits,
first introduced in \cite{Heisenberg}.
Also, formulae which bear some resemblance
to ours appear in \cite{Memb_Matstr}, where a correspondence
between membrane theory and Matrix String Theory is considered.
%%%%%%%%%%%%%%%%%%%%%%%%%v3%%%%%%%%%%%%%%%%%%%%%%%%%%%%%
See also \cite{Taylor}.
%%%%%%%%%%%%%%%%%%%%%%%%%%%%%%%%%%%%%%%%%%%%%%%%%%%%%%%%
}
\beq
\hat{g}_{mn}=g_{m-n}\!\biggl(\!\frac{\hat{f}_m+\hat{f}_n}{2}\!\biggr).
\label{MatrixElement}
\eeq
We also require that when $|m-n|$ gets larger, the value of 
$g_{mn}$ falls off rapidly. This condition naturally conforms 
with (\ref{MatrixElement}), provided that the function $g$ is 
sufficiently smooth and
$N$ is sufficiently large.

A comment to the rule (\ref{MatrixElement}) is in order.
We have freedom to change the orbit $\ss(t)$ by translation of $t$.
The amount of translation is a function of $f$, which we denote by 
$\Delta t(f)$. By this transformation, $g_s(f)$ becomes
\beq
e^{i s \frac{2\pi}{T}\ \!\!\!\Delta\! t(\!f\!)} \ g_s(f).
\eeq
Therefore, $\hat{g}_{mn}$ changes into
\beq
e^{i (m-n) \frac{2\pi}{T}\ \!\!\Delta\! t
\bigl(\!\frac{\hat{f}_m+\hat{f}_n}{2}\!\bigr)} 
\ \hat{g}_{mn}. \label{TimeTranslation}
\eeq
This freedom has a counterpart in the world of matrices.  %v3
Namely, we can change $m$-th eigenvector by a phase factor $e^{i\delta_m}$.
By this transformation, $\hat{g}_{mn}$ becomes
\beq
e^{i (\delta_n -\delta_m)} \hat{g}_{mn}. \label{PhaseChose}
\eeq
Comparing (\ref{TimeTranslation}) and (\ref{PhaseChose}), we find that if
\beq
\delta_n -\delta_m \approx (m-n)\ \frac{2\pi}{T}\ 
\Delta t\ \!\!\bigg(\!\frac{\hat{f}_m+\hat{f}_n}{2}\!\bigg)
\label{RelationBetweenTimetranslationAndPhase}
\eeq
holds, then the two transformations are approximately identical.
We can construct $\delta_m$ satisfying 
(\ref{RelationBetweenTimetranslationAndPhase}) from given
$\Delta t(f)$, provided that $\Delta t(f)$ is sufficiently smooth
and $N$ is sufficiently large.

Equations (\ref{QtzOfArea}) and (\ref{MatrixElement})
constitute, then, 
our new correspondence rule. We shall now deduce
the fundamental approximations (\ref{AppNrm})-(\ref{AppBra})
from the new rule.

By (\ref{QtzOfArea}), we have divided the $\ss$-space into $N$
domains around orbits along which $f(\ss)$ takes the values 
$\hat{f}_1, \cdots, \hat{f}_N$.
We can evaluate the integral of an arbitrary function $\int g(\ss) d^2\ss$
approximately, 
by summing up the average values of $g(\ss)$ on these loops
multiplied by the areas of each domains.
Since (\ref{DefOfgOfs}) and (\ref{MatrixElement}) 
tell us that the average value of $g$
on the orbit along which $f(\ss(t))=\hat{f}_n$ is $\hat{g}_{nn}$, 
and since each area is equal to
$[\ss]/N$, we obtain
\beq
\int g(\ss) \ d^2\!\ss \approx \sum_{n} g_{nn} \frac{[\ss]}{N},
\label{DerivedAppNrm}
\eeq
which is nothing but (\ref{AppNrm}).

We next consider multiplication of matrices constructed by 
(\ref{MatrixElement})
\beq
(\hat{g}\hat{h})_{mn}
=\sum_l \hat{g}_{ml}\ \hat{h}_{ln}
=\sum_l g_{m-l}\ \!\!\biggl(\!\frac{\hat{f}_m+\hat{f}_l}{2}\!\biggr)\ 
h_{l-n}\ \!\!\biggl(\!\frac{\hat{f}_l+\hat{f}_n}{2}\!\biggr).
\label{(gh)mn_basic}
\eeq
Since $g_{m-l}$ and $g_{l-n}$ fall off rapidly when $|m-l|$ and $|l-n|$ 
are large,
respectively,
the terms in which  $l$ is not far away from 
$m$ or $n$ dominate the summation.
Then, by (\ref{QtzOfArea}), neglecting
higher order terms in $1/N$, we can replace both
$(\hat{f}_m+\hat{f}_l)/2$ and $(\hat{f}_l+\hat{f}_n)/2$ by 
$(\hat{f}_m+\hat{f}_n)/2$. We have, therefore,
\beq
(\hat{g}\hat{h})_{mn}\approx \sum_l g_{m-l}\ \!\!
\biggl(\!\frac{\hat{f}_m+\hat{f}_n}{2}\!\biggr)\ 
h_{l-n}\ \!\!\biggl(\!\frac{\hat{f}_m+\hat{f}_n}{2}\!\biggr)=
(gh)_{m-n}\ \!\!\biggl(\!\frac{\hat{f}_m+\hat{f}_n}{2}\!\biggr)
=\widehat{gh}_{mn},
\eeq
where the second equality is the convolution law of Fourier series.
Thus, the matrix corresponding to the multiplication
of the two functions approximately coincides with
the multiplication of matrices corresponding to the functions.
We have derived (\ref{AppMtp}).

We have just seen that, to the leading order, the multiplication of 
the matrices
is commutative, since $\widehat{gh}=\widehat{hg}$, as a matter of course.
Incorporating one more higher order terms in $1/N$,
we shall evaluate the non-commutativity of the matrices.
Thus, from (\ref{(gh)mn_basic}), we have
\beq
(\hat{g}\hat{h})_{mn} \approx 
\sum_l \L(g_{m-l}\ \!\!\biggl(\!\frac{\hat{f}_m+\hat{f}_n}{2}\!\biggr)\ 
+ \frac{\hat{f}_l-\hat{f}_n}{2} g^\prime_{m-l}\R)
\L(h_{l-n}\ \!\!\biggl(\!\frac{\hat{f}_m+\hat{f}_n}{2}\!\biggr)
+ \frac{\hat{f}_l-\hat{f}_m}{2} h^\prime_{l-n}\R).
\eeq
Here, we set $g_s^\prime(f)=dg_s/df$.
\footnote
{
We choose the orbits in (\ref{DefOfgOfs}) smoothly,
so that
$d g_s / d f$ is well-defined.
}
We have omitted the value of $f$ at which $g^\prime$ or $h^\prime$
is evaluated,
since that does not affect the results to the order we are working.
By (\ref{QtzOfArea}), it follows that
\[
(\hat{g}\hat{h})_{mn} \approx \sum_{u+v=m-n}
\L(g_{u}\!\biggl(\!\frac{\hat{f}_m+\hat{f}_n}{2}\!\biggr)
+  \frac{[\ss]}{N}\frac{1}{T}\biggl(\frac{v}{2}\biggr)g^\prime_{u}\R)
\L(h_{v}\!\biggl(\!\frac{\hat{f}_m+\hat{f}_n}{2}\!\biggr)
+  \frac{[\ss]}{N}\frac{1}{T}\biggl(-\frac{u}{2}\biggr)h^\prime_{v}\R),
\]
where we have introduced new dummy indices $u=m-l, v=l-n$.
Then, finally, we have
\beqa
([\hat{g}, \hat{h}])_{mn}
&\approx& \frac{[\ss]}{N}\frac{1}{T}\frac{1}{2}\L(
\sum_{u+v=m-n} \bigl(-g_u (u h^\prime_v) +  (v g^\prime_u) h_v\bigr)
- \big(g\leftrightarrow h\big) \R)  \non\\
&=&
 \frac{[\ss]}{N} \frac{1}{T} \sum
 \big( -\!(u g_u) h^\prime _v + g^\prime_u(v h_v)\ \big). 
\label{ghCommHighFinal}
\eeqa

In order to understand the relation of the last expression to the function 
$\{ g, h\}(\ss)$, it is instructive to consider
the special case $h=f$. Namely, we consider 
the case in which one of the functions is the reference function.
In that case we have by (\ref{QtzOfArea}),
\beq
[\hat{g},\hat{f}]_{mn}
=\hat{g}_{mn} (\hat{f}_n -\hat{f}_m)
\approx \frac{[\ss]}{N} \frac{1}{T} (n-m) g_{m-n}, \label{gfCommFin}
\eeq
which is the special case of (\ref{ghCommHighFinal}).
On the other hands, 
the Lie brackets between $g$ and $f$ can be  
expressed by a solution of  (\ref{EomOnSs}) 
as,
\beq
\{g, f\}(\ss) = \left( \frac{d}{dt} g\left(\ss\!\left(t
\right)\right) \right) 
\Bigg|_{\ss(t)=\ss} \label{HeisenbergEOMinSS},
\eeq
where the total derivative with respect to
 $t$ is taken at the point where the Lie bracket is calculated.
Then, from the definition of the Fourier component $g_s$,
(\ref{DefOfgOfs}),
we get
\beq
\widehat{ \{g, f\} }_{mn} = i (m-n) \frac{2\pi}{T} g_{m-n} .
\label{BragfFin}
\eeq
Comparing with (\ref{gfCommFin}), we obtain
\beq
\widehat{ \{g, f\} }_{mn} \approx  -i \frac{2 \pi N}{[\ss]} 
[\hat{g},\hat{f}]_{mn}, \label{CommIsLieBraSpecial}
\eeq
the special case of (\ref{AppBra}).

The last expression in (\ref{ghCommHighFinal}) and the
above derivation
of (\ref{CommIsLieBraSpecial}) suggest the natural generalization.
We reinterpret the independent
parameter of the ODE, $t$,
as a function defined locally on the $\ss$-space. Then, from 
(\ref{HeisenbergEOMinSS}), we find 
\beq
\{ t , f\} = \frac{dt}{dt} =1,
\eeq
which means that we can consider that $(t, f)$ as canonically conjugate
variables in terms of the analogous canonical formalism.
It follows that,
\beq
\{ g, h \} = \L(\fracp{g}{t}\R)_f \L(\fracp{h}{f}\R)_t - 
\L(\fracp{g}{f}\R)_t \L(\fracp{h}{t}\R)_f. \label{ghBraAsLieBraBytf}
\eeq
The definition of the Fourier components $g_s$, (\ref{DefOfgOfs}), 
is now interpreted 
as the representation of $g$ as a function of $(t ,f)$
\beq
g(t,f)=\sum_{s=-\infty}^{+\infty}
g_s(f) e^{i \left(\frac{2\pi}{T(f)} s \right) t}, \label{gAsAFunctionOftf}
\eeq
Substituting (\ref{gAsAFunctionOftf}) and the similar formula for $h$ 
into (\ref{ghBraAsLieBraBytf})
we get,
\footnote
{
Technically, that $t$ is defined only locally poses a  problem.
However, we can cope with it easily by introducing
patches on each of which $t$ is well-defined, and considering 
the relation between the patches.
}
\beq
(\{g, h\})_s= \sum_{u+v=s} (i \frac{2\pi}{T} u g_u) h^\prime_v
-  g^\prime_u( i \frac{2\pi}{T} v h_v).
\eeq
(Terms in which $(\partial/\partial f)_t $ acts on $1/T(f)$ cancel out.)
By comparing this expression with (\ref{ghCommHighFinal}),
we finally prove (\ref{AppBra}),
\beq
\widehat{ \{g, h\} }_{mn} \approx -i \frac{2\pi N }{[\ss]}
[\hat{g} ,\hat{h} ]_{mn}.
\eeq

Up to this point, our derivation has been confined to the case
where there is only one eigenvalue sequence. 
The extension to the general 
case where there are several eigenvalue sequences
is easy. Namely,
we apply (\ref{QtzOfArea}) and (\ref{MatrixElement}) within
each sequences
separately. 
They determine the
matrix elements  between eigenvectors belonging to 
the same sequence.
We then set remaining matrix elements, that is, 
matrix elements between eigenvectors which belong to 
different sequences, to zero.
Above derivations of the fundamental approximations work just the same.

This argument means that we can concentrate on the
behaviour of functions on one branch of the $\ss$-space, 
ignoring the behaviour on other branches.
Also, since in our arguments
the matrix element $\hat{g}_{mn}$ falls off rapidly
when $|m-n|$ gets larger, we can ignore the behaviour of the functions
at the place differing much in the value of $f$.
These properties render our new rule a semi-local nature.
That is, both the rule and the approximations work 
locally in the direction $f$ changes.
This situation is somewhat reminiscent of the uncertainty principle
in the analogous quantum mechanical case. We have chosen
the representation to make $\hat{f}$ diagonal. This choice 
achieves minimum uncertainty in $f$, 
and, at the same time, makes the conjugate variable $t$
maximally uncertain.
That our rule can be applied to any topology may be 
considered as a direct consequence of this semi-locality.

The linear approximation (\ref{QtzOfArea})
has been essential in the machinery of the derivations
of (\ref{AppNrm})-(\ref{AppBra}).
Therefore, it seems that the linear approximation, hence
the existence of the eigenvalue sequences
is necessary in order that the approximations are good.
Also the use of the analog of the
Bohr-Sommerfeld condition in section \ref{STop} 
is justified, since the condition is nothing but (\ref{QtzOfArea}).

Unfortunately, it seems that our new rule does not 
apply to the following exceptional quantities: matrix elements 
near branching and merging processes.
Our rule is essentially a WKB approximation.
In the immediate 
vicinity of the branching and merging processes, 
there should be 
tunneling effects which make the WKB approximation unreliable.
Consider an analog problem in quantum mechanics, that is,
the motion of a particle in the double-well potential.
It is possible to deal with each well
separately semi-classically,
for sufficiently small $\hbar$, and for generic energy levels.
Indeed, tunneling amplitudes between the wells in general 
are negligibly small,
behaving like $\exp{(-O(1)/\hbar)}$.
However, for those rare energy levels which have energy close to 
the value of the potential at the local maximum,
the tunneling amplitudes are 
not negligible. The break down of our rule could also be expected 
from a more direct argument.
The solution of (\ref{EomOnSs}), $\ss(t)$, in the vicinity of
the branching and merging processes,
spends most of the time near the branching point, moving very slowly.
Then, even if $g(\ss)$ is a smooth function, $g(\ss(t))$ might develop
singularity. Then, the validity of the condition used in our argument, 
that $\hat{g}_{mn}$ is negligible for large $|m-n|$, might be
questioned.

\section{Examples}
\label{SNumEx}
In this section we shall present three examples.
In the first example, by an analytical calculation,
we show the equivalence between our new rule
and the previously known rules. Both
diagonal and 
off-diagonal matrix elements are compared.
In the remaining two examples,
our purpose is mainly to 
illustrate the notion of eigenvalue sequences.
We calculate numerically eigenvalues
of matrices constructed by the previously known rules.
We represent the resulting eigenvalue distribution
in a method such that the structure discussed in section 3,
namely the eigenvalue sequences and their
branching phenomenon, can be easily seen.
We confirm that the branching phenomenon of the
eigenvalue sequences coincides with
that of the orbits of the ODE (\ref{EomOnSs}).
We further numerically compute the eigenvalues
by our rule (\ref{QtzOfArea}), and compare them
 with those calculated by the previously known rules.
\paragraph{Example 1}  \label{ExSphA}
\begin{figure}
\parbox{.02\linewidth}{\ }
\begin{minipage}{.45\linewidth}
\begin{center}
\includegraphics[width=.8\linewidth]{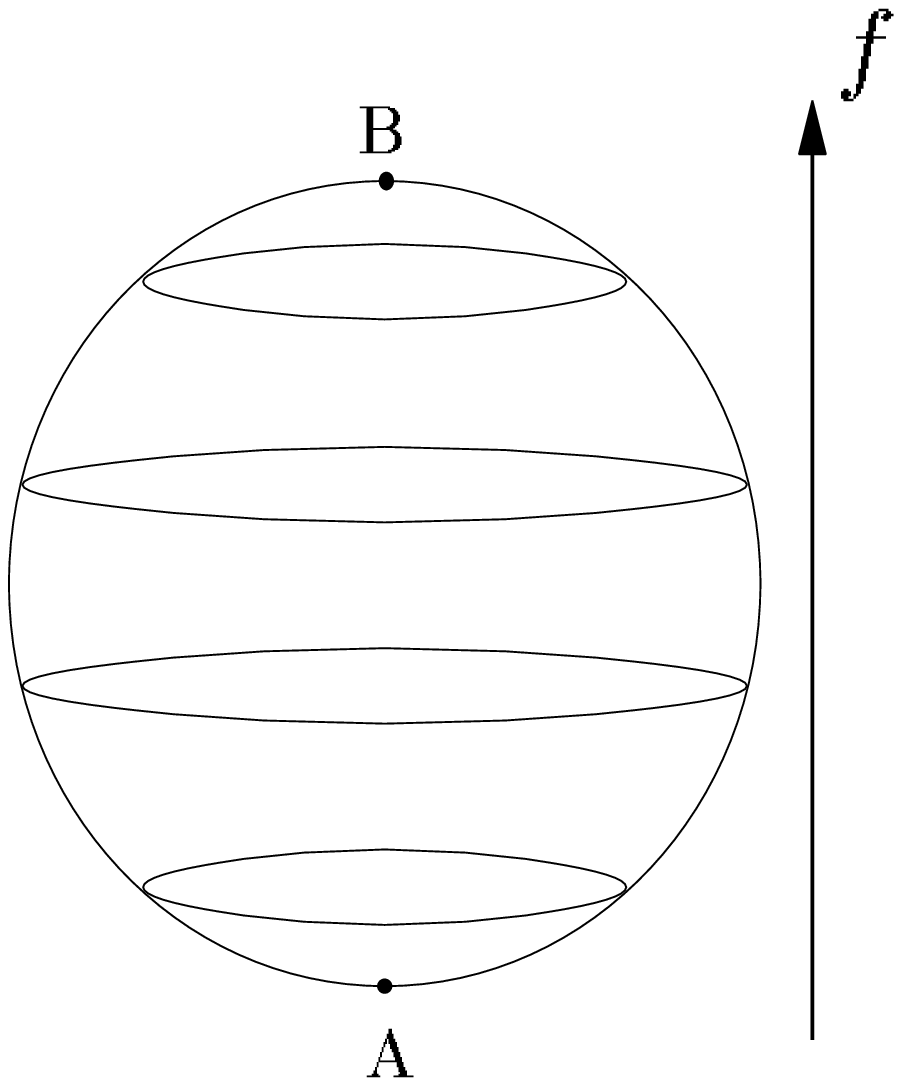}
\caption{The $\ss$-space of spherical topology.
The height is the reference function. The orbits
of (\protect\ref{EomOnSs}) appear at A and disappear at B.
} \label{FSphereA}
\end{center}
\end{minipage}
\parbox{.05\linewidth}{\ }
\begin{minipage}{.45\linewidth}
\begin{center}
\includegraphics[width=.8\linewidth]{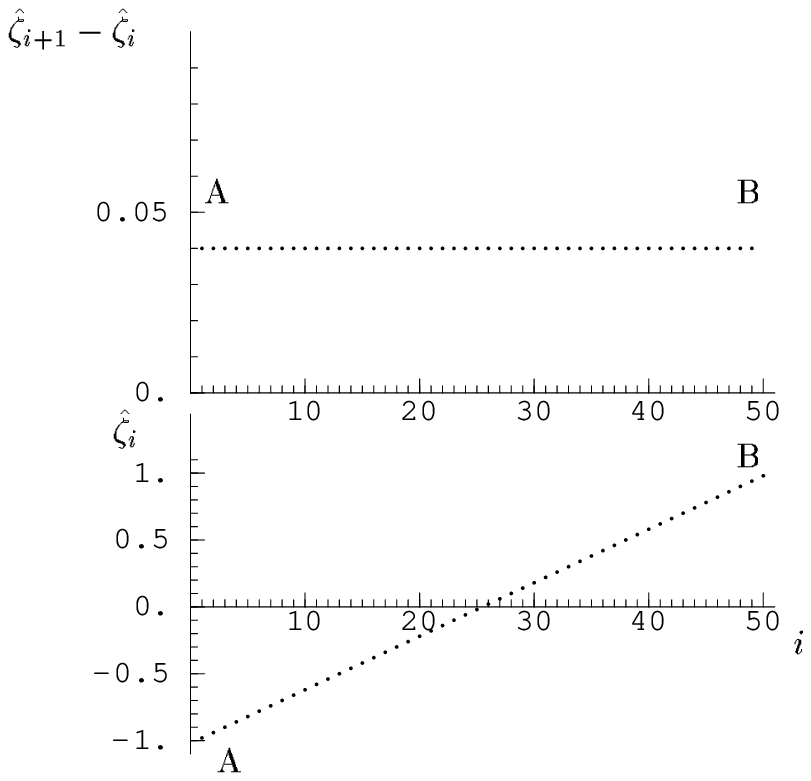}
\caption{Plot of the eigenvalues and the difference
of the eigenvalues of the matrix corresponding
to the height in Fig.~\protect\ref{FSphereA}. The eigenvalue
sequence appears at A and disappears at B.} \label{FSphereAEv}
\end{center}
\end{minipage}
\parbox{.07\linewidth}{\ }
\end{figure} 
We consider the $\ss$-space which has topology of a sphere.
We represent the $\ss$-space as an unit sphere in $\xi, \eta, \zeta$-space,
\beq
\xi^2+\eta^2+\zeta^2=1
\eeq
with the area element given by
\beq
dS = \sin{\theta} d\theta d\phi,
\eeq
where $\theta$ and $\phi$ are polar coordinates defined by 
$\zeta=\cos{\theta}, \xi=\sin{\theta} \cos{\phi},
\eta=\sin{\theta} \sin{\phi}$.
Then the Lie brackets are $\{\xi, \eta\} = \zeta, \cdots$.
We choose the simple reference function $f=\zeta$.
Fig.~\ref{FSphereA} represents the $\ss$-space and the 
reference function.
The orbits of (\ref{EomOnSs})
appear at the point A and disappear at the
point B.

We first construct the matrix $\hat{\zeta}$ corresponding
to the function $\zeta$, by our new rule.
The area of the domain $\zeta \leq \zeta^\prime$ is given by
\beq
\frac{\zeta^\prime+1}{2} 4 \pi.
\eeq
Then, by the analog of the Bohr-Sommerfeld quantization condition,
or (\ref{QtzOfArea}), we obtain
\footnote{
See also footnote \ref{FootnoteOnShiftZeroPoint}.
}
\beq
(\hat{\zeta}_1, \cdots, \hat{\zeta}_N) =
(-1+\frac{1}{N}, -1 + \frac{3}{N}, \cdots, 1-\frac{1}{N}). 
\label{ZetaMatElementsNewRule}
\eeq
We further construct the matrices $\hat{\xi}$ and $\hat{\eta}$,
corresponding to the functions $\xi$ and $\eta$.
The solutions to the ODE (\ref{EomOnSs}) can be explicitly written as, 
\beqa
(\xi+ i \eta)(t) &=&  \sqrt{1-\zeta^2} e^{it}\\
(\xi- i \eta)(t) &=&  \sqrt{1-\zeta^2} e^{-it}.\non
\eeqa
Then, from (\ref{MatrixElement}), the only non-zero
matrix elements is,
\beq
(\hat{\xi} +i \hat{\eta})_{m+1,m}=
\sqrt{1- \L(\frac{\hat{\zeta}_m+\hat{\zeta}_n}{2}\R)^2}
=\sqrt{1- \frac{4}{N^2-1} \L(m-\frac{N}{2}\R)^2}
= (\hat{\xi} -i \hat{\eta})_{m,m+1}. \label{XiEtaMatElmntsNewRule}
\eeq

We shall now compare these results 
with those obtained from the previously known rules.
The rule for the spherical topology reads~\cite{%
Memb_Mr_mrG, Memb_Mr_mrH, S_Memb_Mr},
\beq
\hat{\xi}= \sqrt{\frac{4}{N^2-1}}\hat{l}_x ,\ \ 
\hat{\eta}= \sqrt{\frac{4}{N^2-1}}\hat{l}_y,\ \ 
\hat{\zeta}= \sqrt{\frac{4}{N^2-1}}\hat{l}_z, \label{BasisToBasisSphere}
\eeq
where $\hat{l}_x, \hat{l}_y, \hat{l}_z$ are generators 
of the representation of $SU(2)$ with spin $l=(N-1)/2$.
Since eigenvalues of $\hat{l}_z$ are
$\{-l, -l+1, \cdots, l \}$, we have,
\beq
\L(\hat{\zeta}_1, \cdots, \hat{\zeta}_N\R)=
\L(-\sqrt{\frac{4}{N^2-1}}\frac{N-1}{2}, 
-\sqrt{\frac{4}{N^2-1}}\frac{N+1}{2}, 
\cdots,\sqrt{\frac{4}{N^2-1}}\frac{N-1}{2}\R),
\eeq
which coincides, for large $N$, with the result of our new rule,
(\ref{ZetaMatElementsNewRule}).
Further, it is well known that
in the basis where $\hat{l}_z$ is 
diagonalized, $\hat{l}_x$ and $\hat{l}_y$ have matrix elements only 
between the eigenvectors corresponding to adjacent eigenvalues.
The expression for the non-zero matrix elements are,
\beq
<l_z^\prime+1|(\hat{l}_x+ i\hat{l}_y)|l_z^\prime> 
= \sqrt{l(l+1)-l_z^\prime(l_z^\prime+1)}
= <l_z^\prime|(\hat{l}_x- i\hat{l}_y)|l_z^\prime+1>
\eeq
where we have denoted by $|l_z^\prime>$ the eigenvectors of $\hat{l}_z$ 
belonging to the eigenvalue $l_z^\prime$.
Thus, the result of the previously known rule is,
\beq
(\hat{\xi} +i \hat{\eta})_{m+1,m}=
\sqrt{1- \frac{4}{N^2-1} \L(m-\frac{N-1}{2}\R)\L(m-\frac{N+1}{2}\R)}
=(\hat{\xi} -i \hat{\eta})_{m,m+1}.
\label{SphMatElmntsOldRule}
\eeq
These matrix elements are also approximately equal to 
(\ref{XiEtaMatElmntsNewRule}).
The agreements of our rule with the
previously known rule for the simple functions
$\xi, \eta, \zeta$
imply agreements for more general functions
which can be constructed by multiplying
$\xi, \eta, \zeta$ finite (much less than $N$) times.
The  reason for this is that  
the approximate equality between
multiplication of functions and that of matrices,
(\ref{AppMtp}), is valid for both rules.
\footnote
{
For torus topology, similar argument as
in this example,
using simple functions such as $\cos{\sigma^1}$ or $\sin{\sigma^2}$
(see example 3 for definitions),
has a tricky aspect
since these simple functions are degenerate functions in the sense of 
Morse theory.
}

The eigenvalues $\hat{\zeta}_i$ and the difference 
$\hat{\zeta}_{i+1}-\hat{\zeta}_i$
of the eigenvalues are given in Fig.~\ref{FSphereAEv}.
We see that the eigenvalues consist of one eigenvalue sequence.
The sequence appears at the point A and disappears at the point B
in Fig.~\ref{FSphereAEv}.
They correspond to the branching points of
orbits A, B in Fig.~\ref{FSphereA}.

\paragraph{Example 2} \label{ExSphB}
%%%%%%%%%%%%%%%%%%%%%%%%%%%%%%%%%%%%%%%%%%%%%%%%%%%%%%%%%%%%%
\begin{figure}
\begin{center}
\includegraphics[width=.4\linewidth]{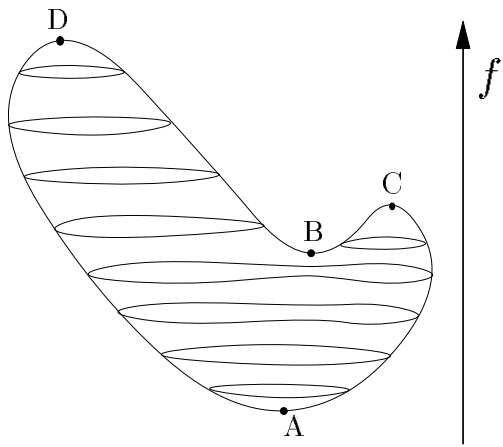}
\end{center}
\caption{
The $\ss$-space of spherical topology 
with a more
interesting branching phenomenon of orbits than
Fig.~\protect\ref{FSphereA}.
Appearing, branching, (first) disappearing and (final) disappearing
processes occur at the points A, B, C and D, respectively. 
} \label{FSphereB}
\end{figure} 
%%%%%%%%%%%%%%%%%%%%%%%%%%%%%%%%%%%%%%%%%%%%%%%%%
%%%%%%%%%%%%%%%%%%%%%%%%%%%%%%%%%%%%%%%%%%%%%%%%%%%%
\begin{figure}
\begin{center}
\includegraphics[width=.6\linewidth]{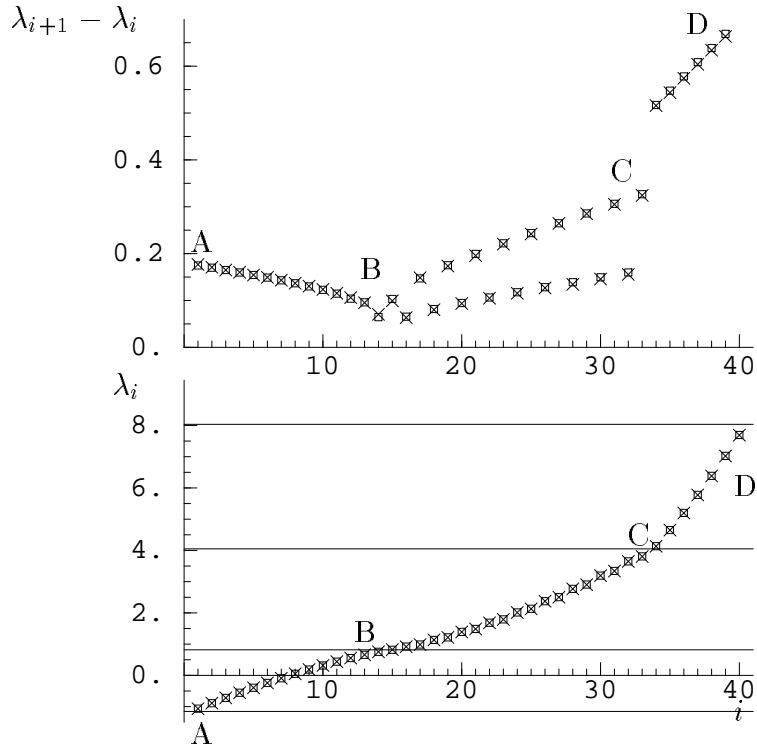}\label{FSphereBEv}
\end{center}
\caption{The plot of the eigenvalues and their difference,
of the matrix corresponding to the reference function
(\protect\ref{RefFuncSphB})
given in
Fig.~\ref{FSphereB}. The eigenvalue distribution
is calculated both by the previously known rule (open squares),
and our new rule, namely, the analog of the Bohr-Sommerfeld condition
(crosses). They agree almost completely.
The branching phenomenon for 
eigenvalue sequences is the same as that of the orbits
in Fig.~\ref{FSphereB}.
Horizontal lines signify critical values of $f$ at which the 
processes in the branching phenomenon take place,
calculated directly from $f$.
}
\end{figure} 
%%%%%%%%%%%%%%%%%%%%%%%%%%%%%%%%%%%%%%%%%%%%%%%%%%%%%%%%%%
We treat another case of spherical topology,
which exhibits a more
interesting branching phenomenon of eigenvalue sequences
than the previous example.
Perturbing the reference function considered there,
we here consider the reference function of the form
\beq
f(\ss)=a\zeta +b\xi + c \xi^2. \label{RefFuncSphB}
\eeq
The reference function
and the $\ss$-space are schematically depicted in Fig.~\ref{FSphereB}.
Orbits of (\ref{EomOnSs})
appear at the point A and then branch into two
families at the point B. Then, the orbits belonging to the right branch
disappear at the point C, and finally the orbits belonging to the
left branch disappear at the point D.

The corresponding matrix $\hat{f}$ is given by 
\beq
\hat{f}=a \sqrt{\frac{4}{N^2-1}} \hat{l}_z + b \sqrt{\frac{4}{N^2-1}} \hat{l}_x+
c\left(\sqrt{\frac{4}{N^2-1}} \hat{l}_x\right)^2, 
\eeq
if one uses the previously known correspondence
rule (\ref{BasisToBasisSphere}). 
We have computed numerically its eigenvalues, in the case
$a=1, b=2, c=6$, with $N=40$.
We have also obtained the eigenvalue distribution
from our new rule.
To this end, we have  computed numerically 
the area of the $\ss$-space as a function of the
height $f$ for each branches of the $\ss$-space.
Then, by the analog of the
Bohr-Sommerfeld condition (\ref{QtzOfArea}),
\footnote
{
See also footnote \ref{FootnoteOnShiftZeroPoint}.
}
we have calculated the eigenvalues of $\hat{f}$.

We represent the eigenvalues by the following method
to see the information of membrane topology.
Firstly, we sort the eigenvalues in increasing order,
\beq
\lambda_1 \le \lambda_2 \le \cdots \le \lambda_N.
\eeq
They are given in Fig.~\ref{FSphereBEv}.
In order to see the branching phenomenon clearly,
it is useful to plot also the difference of the eigenvalues 
$\lambda_{i+1}-\lambda_i$. 
By the plot
%, or just by 
%scrutinising the plot of $\lambda_i$ very closely,
one  finds that
from the point B to the point C,
the plot of $\lambda_i$ is zig-zag shaped.
Thus, the plot of $\lambda_i$ gives a juxtaposition of
four eigenvalue sequences.
We see the same branching phenomenon
of the sequences 
as that of the orbits in Fig.~\ref{FSphereB}.
%We thus have a confirmation of the 
%discussion in section \ref{STop}.
The agreement between our new rule and the previously known
rule is remarkable.

\paragraph{Example 3} \label{ExTorus}
We consider the $\ss$-space which has  topology of a torus. 
The $\ss$-space can be represented by $[0, 2\pi)\times[0, 2\pi)$,
where periodic boundary conditions are understood.
We choose the reference function to be
\beq
f(\ss)=a \cos{\sigma^1} + b \cos{\sigma^2}.
\label{RefFuncTorus}
\eeq
We assume that $a\neq b$, $a\neq 0$, $b\neq 0$, in order to avoid
degenerate reference functions. To be specific we choose $0<a<b$.
%%%%%%%%%%%%%%%%%%%%%%%%%%%%%%%%%%%%%%%%%%%%%%%%%
\begin{figure}
\parbox{0.075\textwidth}{\ }
\parbox[t]{0.4\textwidth}{
(a)
\begin{center}
\includegraphics[height=4.5cm]{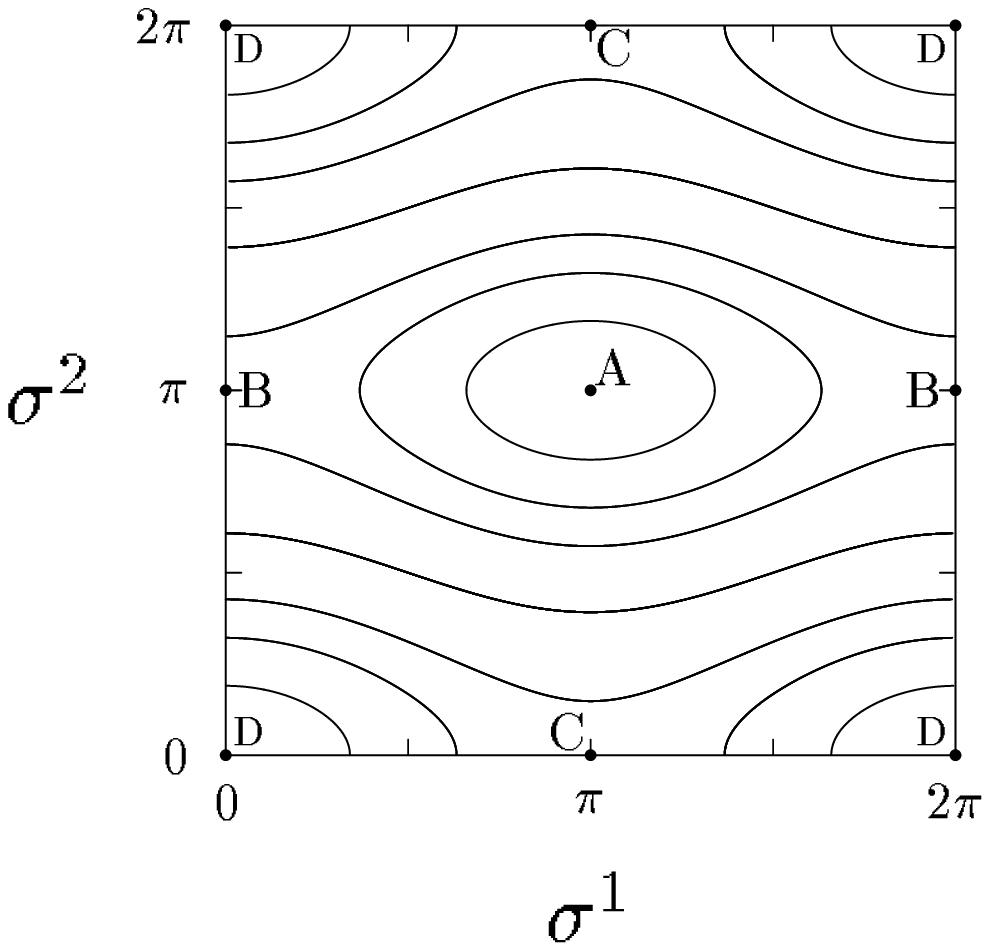}
\end{center}
}
\parbox{0.05\textwidth}{\ }
\parbox[t]{0.4\textwidth}{
(b)
\begin{center}
\includegraphics[height=5cm]{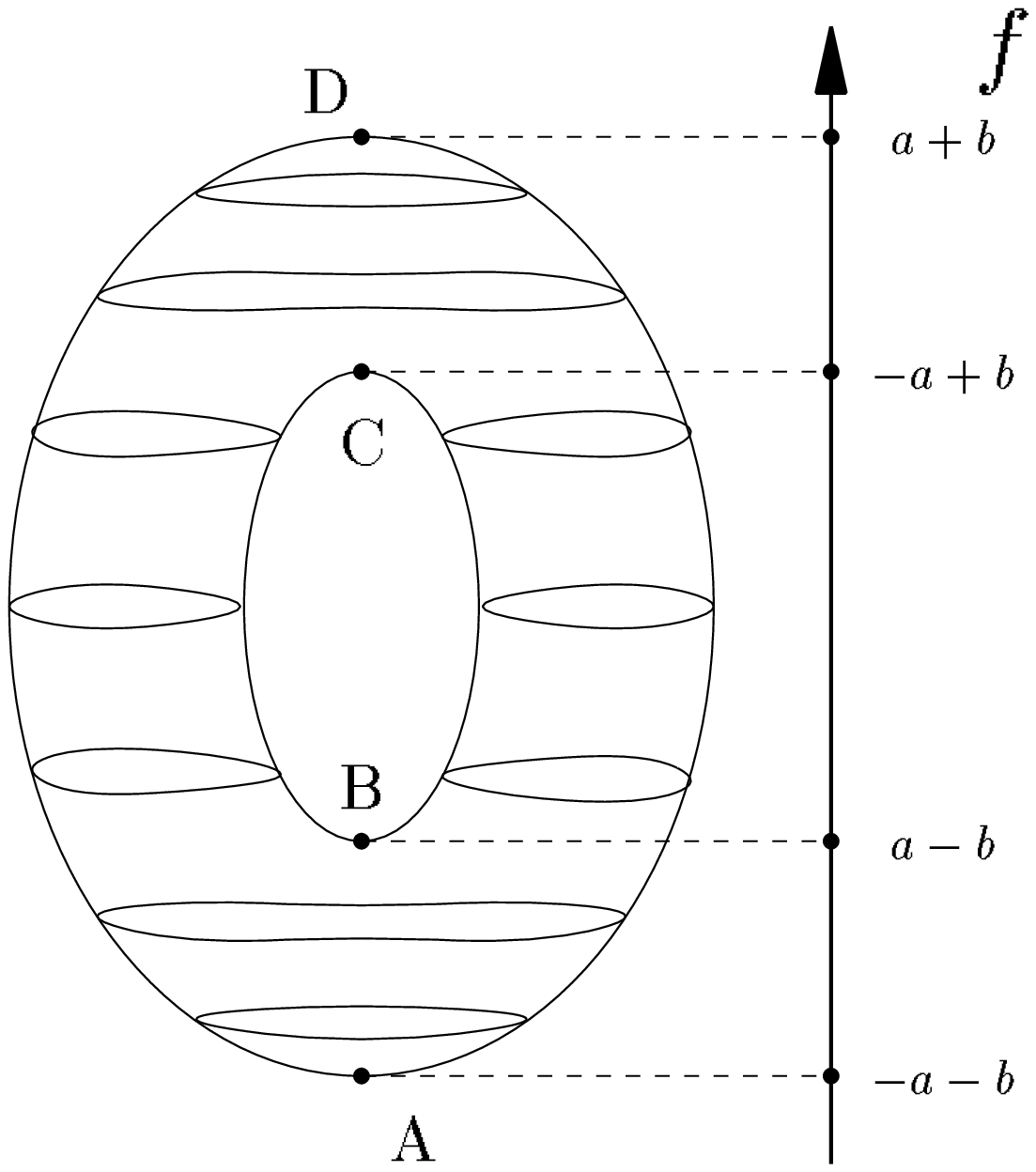}
\end{center}
}
\caption{
(a) Contour plot of the reference function 
$f=a \cos{\sigma^1}+b \cos{\sigma^2}$, with $0<a<b$.
(b) Schematic picture of the $\ss$-space and the reference
function.
The contours, {\it i.e.},
the orbits of (\protect\ref{EomOnSs}) appear, branch, merge and disappear
at the points A, B, C, and D, respectively. 
In (a) the orbits are so written that the areas of the domains
between two adjacent orbits are [\ss]/N.
}
\label{FTorusContour}
\end{figure}
%%%%%%%%%%%%%%%%%%%%%%%%%%%%%%%%%%%%%%%%%%%%%%%%%%%%%%
%
The reference function is represented in Fig.~\ref{FTorusContour}.
It has essentially the same feature
as the reference function in Fig.~\ref{Ftorus}.
At the points A, B, C, D the function $f$ takes the critical values
$-a-b, a-b, -a+b, a+b$, respectively.

\begin{figure}
\begin{center}
\includegraphics[width=.6\linewidth]{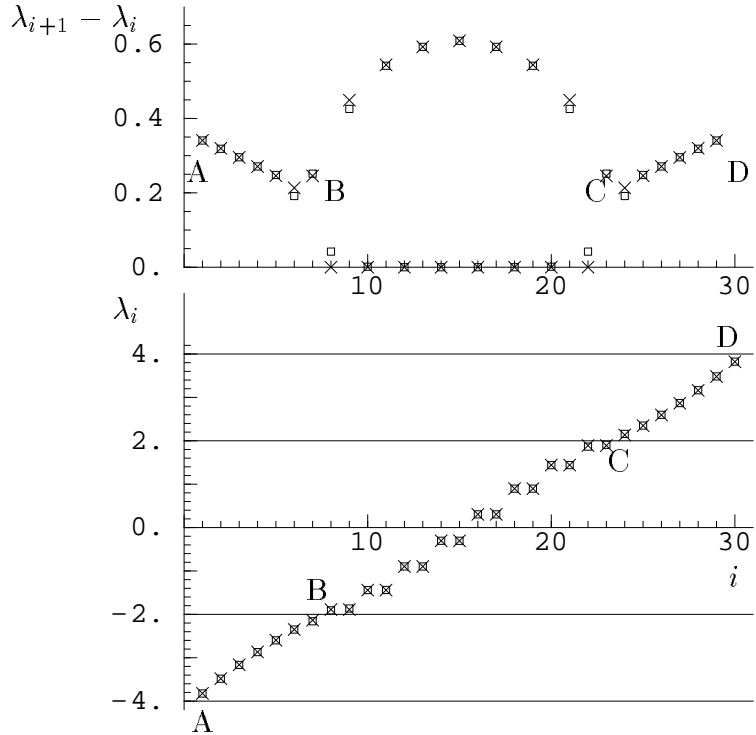}
\end{center}
\caption{
Same as Fig.~\protect\ref{FSphereBEv}
but for the reference function
(\protect\ref{RefFuncTorus}) given in
Fig.~\protect\ref{FTorusContour}.
The topology of the $\ss$-space is that of a torus.
The eigenvalues calculated by the new rule and the previously known
rule agree well except at the immediate
vicinity of the point B or the point C.
The branching phenomenon of  the 
eigenvalue sequences is the same as that of the orbits in 
Fig.~\protect\ref{FTorusContour}.
The (approximate) degeneracy of eigenvalues 
from the point B to the point C is only accidental, 
being result of the symmetry of the $f$.
}
\label{FTorusEigenvalues}
\end{figure}

In the previously known correspondence rule for torus topology,
one postulates\cite{MrTor}, 
\beq
\widehat{e^{i\sigma^1}} = h_1, \ \ \widehat{e^{i\sigma^2}} = h_2, \label{BasisToBasisTorus}
\eeq
where $h_1$ and $h_2$ are the well-known
$N \times N$ matrices which satisfy the relation
$h_1 h_2 =h_2 h_1 \exp{( i 2\pi /N)}$.
Then, it follows that
\beqa
\hat{f}= \frac{a}{2}(h_1+h_1^\dagger)+ \frac{b}{2} (h_2+h_2^\dagger).
\eeqa
We have computed the eigenvalue distribution of this matrix numerically, 
in the case $a=1, b=3,$  with $N=30$.
We can also calculate them by the new rule
as we have done in the previous example. %\ref{ExSphB}.
\footnote{
Due to the poor knowledge in the vicinity of 
the merging and branching processes discussed at
the end of section \ref{SMRCorr},
we have two (or rather one due to the symmetry of the present example)
undetermined parameters of order $1/N$ mentioned in footnote
\ref{FootnoteOnShiftZeroPoint}. 
%Complete fixing of the order $1/N$ parameters requires
%incorporation of the tunneling effect discussed at the end of
%section \ref{SMRCorr}, and beyond the scope of the present paper.
We have fixed the  order $1/N$ parameter by comparison to the result of
the previously known rule.
}
The results by the two methods are given in  Fig.~\ref{FTorusEigenvalues}.
They
 agree well, except at the vicinity of the
branching process at the point B. 
The reason for the discrepancy is 
 noted at the end of the previous section:
our new rule should not be trusted in the vicinity of branching processes.
We can trust the previously known rule, on the other hand,
since the fundamental approximations (\ref{AppNrm})-(\ref{AppBra}) are 
guaranteed by the rule (\ref{BasisToBasisTorus}),
irrespectively of the branching phenomenon.
%%%%%%%%%%%%%%%%%%%%%%%%%%%%%%%%%%%%%%%%%%%%%%%%%%%%%%%%%%%%%%5
%%%%%%%%%%%%%%%%%%%%%%%%%%%%%%%%%%%%%%%%%%%%%%%%%%

In Fig.~\ref{FTorusEigenvalues},
we see a sequence, which appears at the point A and branches
into two sequences at the point B.
Then, the two sequences merge at
the point C,
and finally the last sequence disappears at the point D.
These branching processes of sequences 
directly correspond to the branching processes
of the orbits A, B, C, D in  Fig.~\ref{FTorusContour}.

\section{Conclusions and discussions} \label{SConDis}
In this paper, we have clarified some elementary
but unknown features of the matrix regularization procedure.
We have worked under the simple view that it is an approximation of 
a continuum theory by a discrete theory.
The approximation between two theories is
based solely
on the fundamental approximation formulae (\ref{AppNrm})-(\ref{AppBra}).
We have constructed a new geometrical correspondence rule
between matrices and functions. 
We have shown the validity of the rule directly
by deriving the fundamental approximations from it.
%In the process of the derivation,
%we have also seen what conditions are necessary for 
%the approximations to be good.
%
The new rule is semi-local in the $\ss$-space,
and, as a consequence, can be applied 
to all membrane topologies in a unified way, in marked contrast 
with previously known rules.
Using our rule, 
one can construct functions corresponding to given matrices 
such that the fundamental approximations hold well,
provided that these functions exist.
Whether these functions exist for given matrices can be also determined.
As a physical application, for given matrices
$\hat{x}^\alpha, \hat{p}^\alpha$ of the
matrix model, 
one can construct the geometrical shape and the momentum densities
of the membranes.

The new rule includes
the linear approximation (\ref{QtzOfArea}), 
which is the analog of the Bohr-Sommerfeld condition.
The linear approximation
has lead us to the particular structure
of the eigenvalue distribution,
namely the branching phenomenon of the eigenvalue sequences.
The eigenvalue sequences, which we have introduced in this paper,
are subsets
of the all eigenvalues whose members
can be linearly approximated locally.
From the analog of the Bohr-Sommerfeld condition,
we have shown that
the branching phenomenon reflects the information of topology. 

\begin{figure}
\begin{minipage}{.45\linewidth}
\includegraphics[width=.8\linewidth]{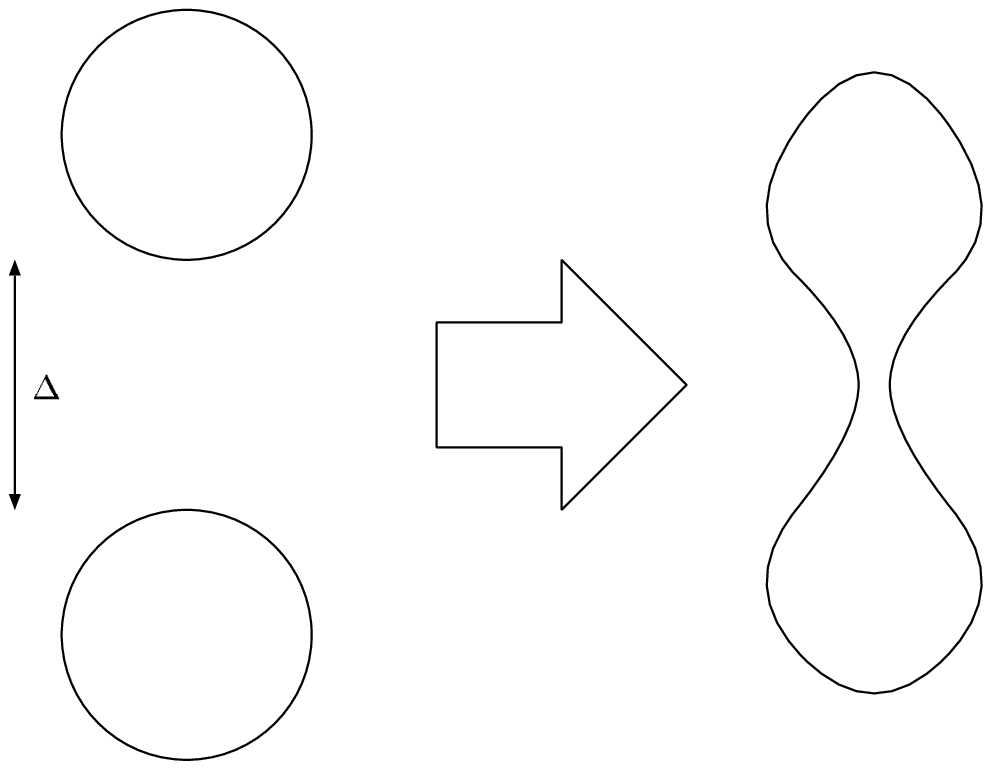}
\caption{Typical case of topology changing of membrane}
\label{FTopologyChangeConti}
\end{minipage}
\parbox{0.1\textwidth}{\ }
\begin{minipage}{.45\linewidth}
\includegraphics[width=.8\linewidth]{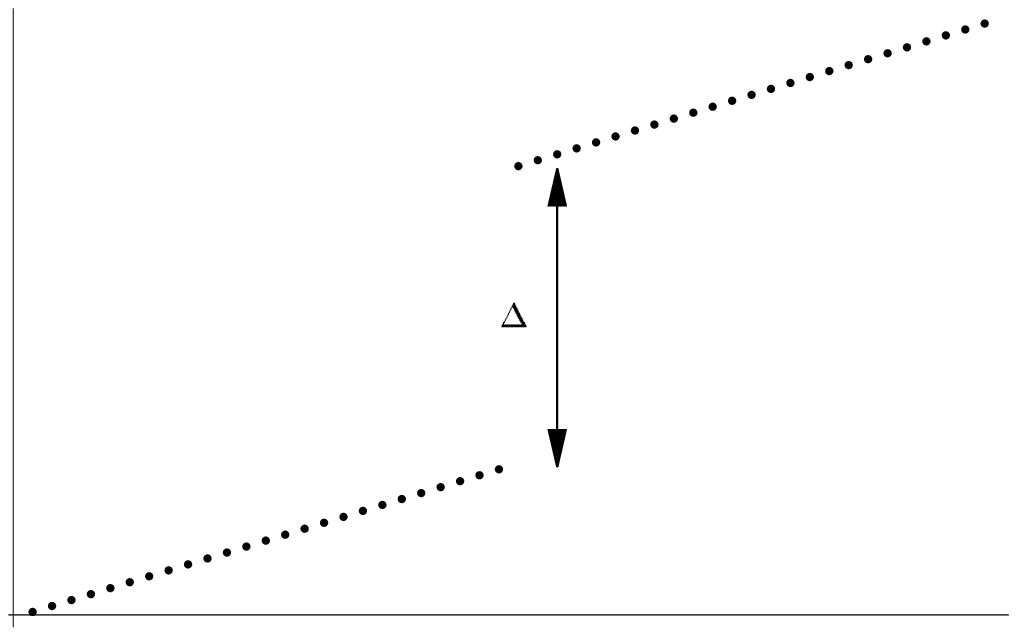}
\caption{Typical case of topology changing from the viewpoint
of eigenvalue distribution}
\label{FTplCngEvals}
\end{minipage}
\end{figure}

Thus, we have clarified the manner
the information of topology
manifests itself
in the eigenvalue distribution.
It is natural to further ask  the question:
``How completely 
can we read off the information of
topology from given matrices?''.
We shall give here some
observations which are essential to this question.
In the first place,
our argument implies that
there is no information of topology in
such ill-behaved matrices for which the fundamental approximation
formulae (\ref{AppNrm})-(\ref{AppBra}) 
do not hold well. 
Indeed, 
it is only for the case (\ref{AppNrm})-(\ref{AppBra}) work,
that the linear approximation
(\ref{QtzOfArea}) should hold.
Hence, even the existence of the eigenvalue sequences
is not guaranteed for those ill-behaved matrices.
Secondly, 
there occurs overlapping 
of topologies when we consider the interaction of membranes. 
For a typical example, let us consider process
shown in
Fig.~\ref{FTopologyChangeConti}.
At first there are two spheres.
Then, these spheres approach each other
and 
the distance $\Delta$ (in the target space)
between two spheres reduces to zero gradually,
and finally the two spheres merge into a sphere.
This overlapping is also present in the eigenvalue distribution.
The process from the viewpoint of eigenvalue distribution
is as follows.
In Fig.~\ref{FTplCngEvals} the 
eigenvalue distribution of the matrix
corresponding to
the height in Fig.~\ref{FTopologyChangeConti} is shown.
The eigenvalue distribution consists of two
eigenvalue sequences. 
If the distance $\Delta$ between the two eigenvalue sequences
gradually reduces to zero, 
then we cannot distinguish the eigenvalue distribution from 
that of a matrix corresponding to 
one sphere.
It is interesting 
to treat topology changing processes of 
membranes by the matrix model in this way.

Our discussion in this paper has been of purely kinematical nature.
To explore the dynamical implication of our rule is
also clearly important.
For example, 
our consideration has made clear the distinction
between the configurations of matrices
which approximate membranes well and
which do not.
It is interesting to consider whether and how the former 
configurations dominate in the path integral of the
matrix model.

We would like to comment on the issue of
the membrane instability\cite{Memb_Instability}.
Let us consider a configuration of membranes 
which has a spike-like portion whose area is less than $1/N$.
If we simply apply the analog of the Bohr-Sommerfeld
condition, we should fail to include the information of the spike
into the matrices. Stated more appropriately,
our argument tells us that
the configuration
cannot be well approximated by $N\times N$ matrices.
We want to stress that this spike has an essential difference 
to the spike which is considered in  the  membrane instability.
One uses the word spike for a portion of a surface
 when its linear dimension is large,
and at the same time its area is small. The difference between the spike
in our context and the spike in the instability context lies 
in the meaning of the area.
In the former, the area means area in the $\ss$-space, that is 
essentially $p^+$.
In the latter, the area means area in the target space, or
the energy of the spike.
This difference is meaningful. Indeed,
there are membranes which have portions that have 
small energy but large $p^+$ or vice versa.
In particular, one can construct configurations of the matrix model
which approximate well the membranes with spikes in the sense of the
membrane instability. 

As a direction of further investigation of the matrix regularization
procedure itself,
we recall the discussion at the end of section
\ref{SMRCorr}.
Our rule in the present form does not include tunneling effects between
sequences. Although our rule
gives correct overall behaviour of the matrix elements,
we could not trust the rule in the present form
to investigate the matrix elements in the immediate vicinity of 
the merging and branching processes. 
Concrete
examples of the processes are the points B, C in example 3 in section 5.
%the point B in example 2 is also branching
\footnote{
We can trust our rule 
near appearing or disappearing processes,
such as the point C
in example 2 of section \ref{SNumEx}.
The jump
in the plot of the difference of eigenvalues 
is a natural consequence of our rule.
}
It is an important task
to extend our rule to incorporate the tunneling effects\cite{Tunneling}.
One possible strategy would be  to revisit the analog problem,
namely the quantum mechanics of a particle in a double-well potential.
We can construct a formula to relate
the semi-classical wave functions in both wells, which is valid
even for the energy level
near the local maximum, extending the ordinary argument
in the WKB approximation
using Airy functions.
Another interesting question
is the uniqueness
of the correspondence which gives the fundamental approximations.
Although we cannot, at present, provide the proof,
we suspect that the correspondence rule
from which (\ref{AppNrm})-(\ref{AppBra}) can be derived
is unique up to similarity transformation.
Indeed, examples studied in section 5 suggest
that the previously known rules and our new rule are
the same up to similarity transformation.
An immediate consequence of the uniqueness is that
a change of the reference function
should amount to a similarity transformation.
\footnote{
This property implies that
the fundamental approximations are no good
for a configuration of matrices which 
consists of matrices corresponding to different topologies.
}

%%%%%%%%%%%%%%%%%%%%%%%%%%%%%%%%%%%%v3 %%%%%%%%%%%%%%%%%%%%%%%
It is believed that
the matrix regularization can be
extended for general even dimensional base spaces on which the
Lie brackets can be defined.
The correspondence rule between matrices and functions
can be easily constructed by using tensor product of matrices,
when the topology of the base space are given by direct
product of some two dimensional spaces.
These extensions are important in the matrix model
of M-theory, in order to incorporate longitudinal 5-branes.
To extend our analysis to 
study topological properties of these higher dimensional objects
is also an interesting problem.
We believe that the analysis analogous to the WKB approximation,
used throughout in this paper, will also be useful
for the higher dimensional case.
However, it would be a challenging task,
since the WKB approximation itself is not fully understood
for generic non-integrable Hamiltonian systems on four 
or more dimensional phase spaces,
compared to that for the necessarily integrable systems
on two dimensional phase spaces considered in this paper.
%%%%%%%%%%%%%%%%%%%%%%%%%%%%%%%%%%%%%%%%%%%%%%%%%%%%%%%%%%%%%

We conclude 
with three possible applications of the new correspondence rule.

(1) One can construct various 
interesting configurations of the matrix model by our rule. 
A particular merit of our rule
in this respect is that it can be applied to a membrane
which has genus higher than two, with no more difficulty 
than to a membrane having topology of a sphere or a torus.

(2) Our rule may be useful when investigating
the Lorentz symmetry of 
the matrix model.
It has been tried to regularize the Lorentz generators of 
the continuum theory in order to construct those of the matrix model
\cite{MembLorentzSymm}.
However, since some of the Lorentz generators are not built up by
simple multiplications or integrations or Lie brackets,
one has been inclined to use the basis expansion of the
previously known
correspondence rules.
Since there are many different expansions for different topologies,
it has been difficult to define the
Lorentz generators in a unique way.
Since our rule can be applied uniformly to all topologies,
it is a promising tool in 
constructing
definitions of the Lorentz generators of the matrix model in a unique way.

(3) The geometrical interpretation
of matrix elements in our rule may make 
the problem of the large $N$ limit of the matrix model accessible.
The problem can be interpreted as
a renormalization of the membrane theory.
We hope that our rule,
by determining the short-distance (or rather, the small-area)
degrees of freedom,
enables us to construct a block-spin transformation 
of the matrix model.

\paragraph{}
I would like to thank, first of all,
Prof. T. Yoneya for discussions, encouragements
 and careful reading of the manuscript.
I would like to thank, for discussions and encouragements, 
other members and
former members of Komaba particle theory group,
especially Y. Aisaka, S. Dobashi
and T. Sato.
Also, I would like to thank W. Taylor, S. Iso and T. Shimada
for discussions and encouragements, %%%%%%%%%%v3%%%%%
and J. Hoppe for helpful and encouraging correspondence.
%%%%%%%%%%%%%%%%%%%%%%%%%%%% v3%%%%%%%%%%%%%%%%%%%%%%%%%

%%%%%%%%%%%%%%%%%%%%%%%%%%%%%%% v3 %%%%%%%%%%%%%%%%%%%%%%%%%%%%%
\paragraph{Note Added:}
This paper is an extended version of
the author's master thesis \cite{MasterThesis}
submitted to University of Tokyo on April 2002,
where the basic results %in this paper
were preliminarily reported.
The author has recently noticed that Hyakutake
has constructed matrices corresponding to
axial symmetric membrane configurations \cite{Hyakutake},
which are special cases of the general prescription given in 
section \ref{SMRCorr} of this paper.
%%%%%%%%%%%%%%%%%%%%%%%%%%%%%%% v3 %%%%%%%%%%%%%%%%%%%%%%%%%%%%

\end{document}